\documentclass[a4paper,11pt]{article}
\usepackage{jinstpub} % for details on the use of the package, please see the JINST-author-manual
\usepackage{lineno}
\usepackage{subcaption}

\title{Reusable Verification Components for High-Energy Physics readout ASICs}

\author[a,1]{M.~Lupi,\note{Corresponding author.}}
\author[a]{S.~Esposito,}
\author[a]{X.~Llopart-Cudie,}
\author[a]{A.~Pulli,}
\author[a]{S.~Scarf\'i,}
\author[b]{N.~Kharwadkar}

\affiliation[a]{CERN, 1, Esplanade des Particules, Meyrin, Switzerland}
\affiliation[b]{FNAL, Batavia IL 60510, USA}

\emailAdd{matteo.lupi@cern.ch}

\abstract{
Verification is a critical aspect of designing front-end (FE) readout ASICs for High-Energy Physics (HEP) experiments. 
These ASICs share several similar functional features, resulting in similar verification objectives, which can be addressed using comparable verification strategies. 
This contribution presents a set of re-usable verification components for addressing common verification tasks, such as clock generation, reset handling, configuration, as well as hit and fault injections. 
The components were developed as part of the CHIPS initiative and they have been successfully used in the verification of multiple HEP ASICs.
}


\arxivnumber{2411.03478} 

\begin{document}
\maketitle
\flushbottom
\newpage

\section{Introduction}\label{sec:intro}
ASICs for detector readout developed by the High-Energy Physics (HEP) community can be divided into two main categories: front-end (FE) readout and concentrator ASICs.

FE readout ASICs are responsible for receiving analog signals from a sensor, converting them into digital signals, and transmitting the resulting values outside of the ASIC.
Typically, a FE readout ASIC includes multiple readout channels, as well as a data transport layer that combines the data from all active readout channels and sends it to the detector readout.
Each readout channel is a mixed signal circuit.
The digital module carries out functions such as filtering, data compression, and transmission of the resultant values.
FE readout ASICs perform similar functions with minor variations to meet detector requirements.

Concentrator ASICs are responsible for receiving data from multiple FE readout ASICs, processing them, and sending them out on one or more high-speed serial interfaces.
The data processing includes alignment, error detection, zero-suppression, data compression, and re-ordering.

Functional verification of these ASICs is fundamental to implement shift-left strategies to achieve first-time-silicon success.
% ~\cite{siemens:shiftleft}.
Given the similarities between HEP ASICs, it can be beneficial to re-use verification components avoiding duplication of work.
This contribution describes a set of re-usable verification components, developed at CERN under the CHIPS initiative, that allow bringing up a verification environment in hours instead of weeks.

\section{Universal Verification Methodology}\label{sec:uvm}
The Universal Verification Methodology (UVM) is, de facto, the industry standard for functional verification.
Developed at the beginning of the 2010s, it is now an IEEE standard~\cite{ieee:uvm} since 2020.
% https://accellera.org/downloads/standards/uvm

UVM defines a set of base classes which allow building a comprehensive, reusable, and scalable verification environment.
It provides a standardized framework for developing a testbench which leverages advanced verification techniques.
\begin{itemize}
\item Constrained-random stimuli generation: the stimuli driving the design are not predetermined but they are instead randomized, inside a pool of valid stimuli.
\item Functional coverage analysis: one of the big questions of verification is ''When are we done verifying?''.
The answer to this question lies in the functional coverage.
The verification engineer, in accordance with the designer, can define a set of scenarios under which the DUT needs to be exercised: these are then translated by the verification engineer into \emph{Systemverilog covergroups} and \emph{coverpoints}~\cite{ieee:sv:lrm}.
The verification environment then allows gathering coverage from a set of tests to understand when all the predefined scenarios have been observed.
\item Reusable testbench components: UVM encourages the design of hierarchical and reusable UVM verification components (UVCs)~\cite{siemens:uvc}.
Re-usability introduces overhead in early stages but saves time in the long run.
\end{itemize}

UVM is widely supported by the major EDA tools.
In recent years, support has also been added to open-source simulators.

\section{UVM verification components}\label{sec:uvc}
In this contribution, we focus on the description of reusable UVCs.
The contribution is meant to be an overview of the available UVCs, however, it does not replace the integration guide available for each UVC.
The components were developed as part of the verification environment of the ASICs indicated in section~\ref{sec:end}.
They are made available to the HEP community via the CERN ASIC support~\cite{asicsupport} in the form of GitLab repositories.

The UVCs presented here are wrapped in the form of UVM environments or UVM agents.
The components implement an interface allowing the connection to the DUT and a monitor to observe and reconstruct the transactions on the interface.
They also implement sequencer/driver pairs and they are complemented by predefined sequences that the user can extend: these will only be instantiated when the UVC is used as an active component.

\subsection{Pin UVC}\label{sec:uvc:pin}
HEP ASICs are generally designed with several General-Purpose I/O (GPIO) pins that are either static (e.g. device ID) or dynamic (reset, trigger, sync, monitor, etc.). 
Pin UVC implements UVM collaterals to control and monitor these single-ended signals.

To configure this UVC, users need to specify several input parameters: the number of pins to be managed, the initial state of each pin, and whether the assertion and de-assertion of the signal should be synchronous or asynchronous. 

The UVC is provided with sequences that allow users to toggle the signal or generate a single pulse for testing purposes. 
Additionally, it includes randomizable parameters to vary the duration and the high or low time of the pulse, adding flexibility and enhancing the testing scope.

\subsection{Clock UVC}\label{sec:uvc:clk}
The Clock UVC is intended to generate a set of related clock signals for various elements within a verification environment, such as DUTs or other UVCs. 

This component allows users to configure key clock attributes, such as the period of the fastest clock and the ratio between the periods of the slowest and fastest clocks, specified as an integer. 

It also offers versatile sequence capabilities, allowing for the generation of clock signals that are integer multiples of the fastest clock period. 
Additionally, users can opt to generate clocks with or without jitter, adding flexibility for diverse testing scenarios. 

Moreover, the Clock UVC includes randomizable parameters to enable variation in clock generation. 
These parameters include the ratio relative to the fastest clock, maximum jitter, and clock phase, providing robust control for testing needs.

\subsection{I2C UVC}\label{sec:uvc:i2c}
The I2C UVC allows controlling an I2C bus~\cite{nxp:i2c}, enabling various functionalities crucial for test environments. 
It is applicable in several key scenarios: configuring the DUT within both top-level and system-level testbenches and interfacing seamlessly with the UVM Register Access Layer (RAL).
The UVC provides an abstraction layer compatible both with a simplified version of the UVC (\emph{simple mode}) or major EDA vendor VIP (\emph{VIP mode}), although a license is required for the latter.

Configuration options allow users to select between a \emph{simple mode} and \emph{VIP mode}, define the DUT device ID, set the address and data lengths, and specify whether a repeated start or stop condition is required, including a start-on-read option.

This component supports running a full range of I2C sequences on the bus and allows DUT access through the RAL. 
Additionally, certain parameters are randomized, including the device ID, address, data, and access type, allowing for varied and comprehensive test coverage.

\subsection{Wishbone UVC}\label{sec:uvc:wb}
The Wishbone UVC is designed to manage and test Wishbone bus operations effectively within a UVM environment. 
This component is applicable for several primary use cases: configuring the DUT in module-level testbenches, verifying the operations of the Wishbone master, and integrating seamlessly with the UVM Register Abstraction Layer (RAL). 

For configuration, the UVC accepts inputs that allow users to define its operation as either a master or slave on the bus, along with setting address and data lengths. 
The component supports a variety of Wishbone sequences, which can be run directly on the bus, and enables DUT access through the RAL. 
Additionally, the UVC includes randomizable parameters for address, data, and access, enhancing test coverage through variability in simulation runs.

\subsection{Hit UVC}\label{sec:uvc:hit}
The Hit UVC is designed to facilitate the simulation of particle hits within a pixel-like readout chip. 
Its primary use cases include providing a structured methodology for injecting particle hits into front-end readout chips and verifying the operation of these chips under simulated particle interactions.

The component accepts several key user inputs, such as the level of integration effort and the matrix dimensions, defined by $x$-size and $y$-size, which shape the spatial framework for the hit locations. 
It supports sequences for hit injection, either by reading predefined hit patterns from an external file or by generating random hits.

In terms of customization, the UVC offers randomizable parameters, allowing users to vary the hit location (specified in $x$ and $y$ coordinates), the charge associated with each hit, and the timing between successive hits. 
This flexibility enables diverse test scenarios, enhancing the thoroughness of the verification process for the front-end readout chip.

\subsection{SEE UVC}\label{sec:uvc:see}
The SEE UVC is designed to simulate and evaluate the impact of Single Event Effects (SEEs), including both Single Event Upsets (SEUs) and Single Event Transients (SETs), on digital designs. 
This component primarily facilitates three use cases: it provides a structured methodology for injecting SEEs, allows for the exploration of design susceptibility to these events, and enables verification of the system’s functionality under SEE conditions. 

To operate, the UVC requires specific user inputs, including integration scripts, a list of nodes where faults may be introduced, and a strategy for filtering these nodes to focus on relevant areas. 

The UVC follows a defined sequence of operations: initiating the fault injection, stopping it as required, and re-starting the injection if necessary to support different phases of testing. 
Moreover, the UVC offers randomizable parameters, allowing users to vary the fault type (selecting from SET, SEU, or generic SEE), the specific node affected, and the duration of the fault, to create diverse test scenarios and thoroughly examine design robustness.

\section{Integration example}\label{sec:example}
\begin{figure}[thbp]
\begin{subfigure}{.5\textwidth}
  \centering
  \includegraphics[width=\textwidth]{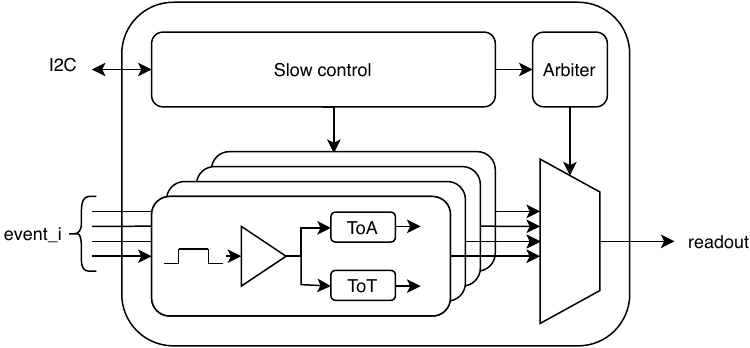} 
  \caption{}\label{fig:tb:dut}
\end{subfigure}%
\begin{subfigure}{.5\textwidth}
  \centering
  \includegraphics[width=\textwidth]{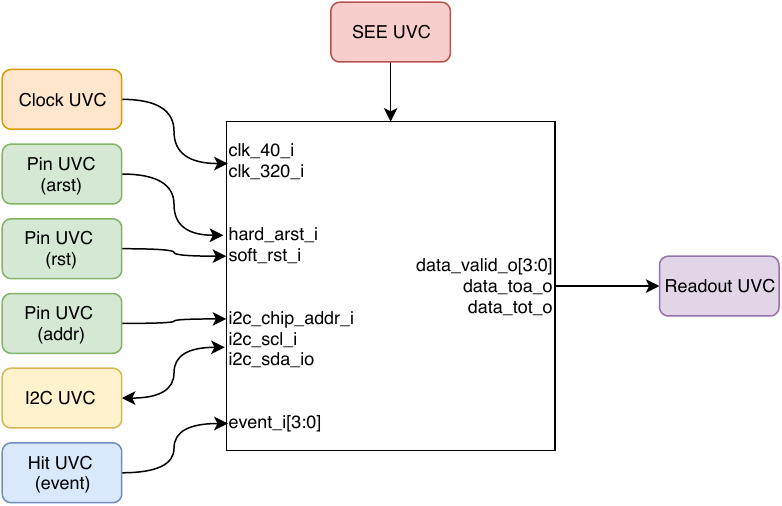}
  \caption{}\label{fig:tb:tb}
\end{subfigure}
\caption{Example of DUT (a) and testbench with the UVCs connected to the 
DUT (b).\label{fig:tb}}
\end{figure}

In this section, we're describing a simple DUT, the \emph{event recorder}, in figure~\ref{fig:tb:dut}, as an example of how the aforementioned UVCs can be used to bring up the verification environment for the DUT.
This example is also available for the community through the CERN ASIC support~\cite{asicsupport}.

The DUT emulates a 4-channels front-end ASIC with Time of Arrival (TOA) and Time over Threshold (TOT) measurement of the input signal \emph{event\_i}.
The DUT receives two synchronous clocks: a slow ($40MHz$) and a fast ($320MHz$) one.
Two different resets are available: an asynchronous hard reset (resetting the whole DUT) and a $40MHz$-synchronous soft reset (not affecting the configuration registers).
The DUT user configuration can be achieved using the I2C protocol.
The device ID on the I2C bus is configured by four dedicated pins, that need to be stable over the operation of the DUT.

The corresponding testbench for the DUT, in figure~\ref{fig:tb:tb}.
Most of the UVCs above can be used to steer the inputs of the DUT:
\begin{itemize}
    \item The clock UVC can be used to provide two synchronous clocks at the required frequencies. 
    The two clocks will be synchronous by design.
    The effect of the input clock jitter can be modeled simply by constraining the maximum jitter in the UVC configuration.
    \item The Pin UVC can be used to control three different interfaces: the two resets and the I2C device ID.
    Two different instances of the Pin UVC will control the resets.
    These can be configured to be toggled asynchronously, for the hard reset, or synchronously to the $40MHz$ clock, for the soft reset.
    The resets can be toggled using the provided sequences.
    An additional instance of the Pin UVC can be used to statically drive the I2C device ID.
    This instance will not make use of sequences to steer the value of the GPIO, but its value can be randomized during the configuration of the UVC.
    \item The I2C bus can be driven using the I2C UVC.
    The device ID of the DUT can be constrained to match the value provided via the Pin UVC.
    The verification engineer can provide a RAL and bind it to the I2C UVC via the provided adapter.
    This will allow testing randomized configurations of the DUT.
    \item In order to inject particle hits data into the DUT, the Hit UVC can be used.
    The user will have to properly constrain the hierarchical path to the node where to inject (force and release) the hit.
    The verification engineer must provide a sequence to inject the single hits.
    This depends on the expected rate of events to be observed by the ASIC.
    The UVC will also provide a transaction item which can be used in the reference model of the DUT.
    \item Finally the SEE UVC can be used, once the verification framework is in place to inject SEEs into the DUT.
    This allows verifying resilience to SEUs in the DUT triplicated RTL, and to SEEs in the gate-level netlist of the implemented DUT.
    Some examples of potential fault injection campaigns are described here~\cite{vrf:sim2023}.
\end{itemize}

The verification engineer is left with the task of designing only the readout UVC and the DUT reference model and scoreboard.

%\begin{figure}[thbp]
%\centering
%\includegraphics[width=.45\textwidth]{img/dut.pdf}
%\qquad    
%\includegraphics[width=.45\textwidth]{img/event_recorder.pdf}
%\caption{Example of DUT (left) and testbench with the UVCs connected to the DUT (right).\label{fig:tb}}
%\end{figure}

%\begin{figure}[thbp]
%\centering
%\includegraphics[width=.6\textwidth]{img/dut.pdf}
%\caption{Example DUT: event recorder\label{fig:dut}}
%\end{figure}
%
%\begin{figure}[thbp]
%\centering
%\includegraphics[width=.6\textwidth]{img/event_recorder.pdf}
%\caption{Example of testbench with the UVCs connected to the DUT\label{fig:tb}}
%\end{figure}

\section{Conclusions}\label{sec:end}
The UVCs presented in this contribution have been successfully used in the verification of multiple ASICs: ATLAS HGTD ALTIROC~\cite{altiroc2023}, ATLAS/CMS RD53~\cite{rd53:atlas}, CMS HGCAL ECON-D and ECON-T~\cite{econ2024}, and EXP28~\cite{exp28}.
Moreover, they are currently being used in the verification of the following ASICs: ALICE ITS MOSAIX~\cite{tdr:its3}, LHCb RICH Fastrich~\cite{tdr:lhcb:rich}, and in the LA-Picopix~\footnote[2]{LA-Picopix: A hybrid pixel detector under development with a $256 \times 256$ pixel particle-tracking detector readout ASIC, achieving <30 ps RMS timing resolution. Features include on-chip clustering, data packet sorting, and a readout rate up to $3.9 \times 10^9$ clusters/s.}.

\end{document}